\documentstyle[aps,preprint]{revtex}
\tightenlines
\begin{document}
\draft
\title{OBSERVATION OF $\rho / \omega$ MESON  MODIFICATION IN NUCLEAR MATTER}
\author{K.~Ozawa\thanks{Present Address: Center for Nuclear Study, 
    Graduate School of Science,  University of Tokyo, 
        7-3-1 Hongo, Tokyo 113-0033, Japan, 
        email: ozawa@cns.s.u-tokyo.ac.jp},
 H.~En'yo, H.~Funahashi, M.~Kitaguchi, M.~Ishino\thanks{Present
   Address: ICEPP, University of Tokyo, 7-3-1 Hongo, Tokyo 113-0033,
   Japan}, 
 H.~Kanda\thanks{Present Address: 
   Physics Department, Graduate School of Science, Tohoku University,
   Sendai 980-8578, Japan},
  S.~Mihara$^\dagger$, T.~Miyashita\thanks{Present Address:
    Fujitsu Corporation, 4-1-1, Kamikodanaka, Nakahara,
    Kawasaki, Kanagawa 211-8588, Japan}, 
  T.~Murakami, R.~Muto, M.~Naruki, 
  F.~Sakuma, H.~D.~Sato, T.~Tabaru, S.~Yamada, 
  S.~Yokkaichi\thanks{Present
   Address: RIKEN, 2-1 Hirosawa, Wako, Saitama 351-0198,
   Japan}
  and ~Y.~Yoshimura\thanks{Present Address: 
    Xaxon Corporation,1-3-19, Tanimachi, Chu-ou, Osaka, Japan}}
\address{Department of Physics, Kyoto University, 
Kitashirakawa Sakyo-Ku, Kyoto 606-8502, Japan }
\author{J.~Chiba, M.~Ieiri, M.~Nomachi\thanks{Present Address: 
    Department of Physics,
     Osaka University, 1-1 Machikaneyama, Toyonaka, Osaka 560-0043
     , Japan}, O.~Sasaki, 
  M.~Sekimoto and  K.H.~Tanaka} 
\address{Institute of Particle and Nuclear Studies, 
        , KEK, 1-1 Oho, Tsukuba, Ibaraki 305-0801, Japan}
\author{H.~Hamagaki}
\address{Center for Nuclear Study, Graduate School of Science,  
  University of Tokyo, 
        7-3-1 Hongo, Tokyo 113-0033, Japan}

\date{\today}
\maketitle
\begin{abstract}
We have  measured invariant mass spectra of electron-positron pairs 
 in the target rapidity region of 12GeV p+A reactions. We have observed a 
significant difference in the mass spectra below the $\omega$  
meson between 
p$+$C and  p$+$Cu interactions. The difference is interpreted
as a signature of the $\rho$/$\omega$ modification at normal 
nuclear-matter density.
\end{abstract}
\pacs{24.85.+p,25.30.-c}

\narrowtext

Recently, chiral property of QCD
in hot(T$\neq$ 0) or dense($\rho\neq$ 0) matter 
attracts wide interests in the field of hadron physics.
The dynamical breaking of chiral symmetry in the QCD
vacuum induces an effective mass of quarks, 
known as constituent quark mass.
In hot and/or dense matter this broken symmetry is 
subject to restore partially or completely,
and hence the properties  of hadrons can be 
modified. To observe such an effect, measurements
of in-medium decay of vector mesons, especially
 in the lepton-pair channel,
 are highly desirable to obtain directly
the meson properties in matter \cite{experiments}.

Although many heavy-ion experiments 
were carried out in  CERN-SPS and BNL-AGS to study hot and dense matter, 
there was only an experiment which could address 
the mass modification of vector mesons. 
The CERES/NA45 collaboration measured low-mass electron 
pair productions in Pb-Au collisions at 158 A GeV
\cite{ceres}, and  observed an  enhancement 
of e$^+$e$^-$ pair yield  in the mass range 
0.3 $<$ $m_{ee}$ $<$ 0.7 GeV/c$^2$  over the expected 
yield from the known hadronic sources in pp  collisions.
This striking effect could be understood as a consequence  
of the mass modification of the $\rho$ and $\omega$ meson
in hot matter.

In  QCD  the mass of  vector mesons,
mainly determined by the effective mass of quarks,
is closely related to $\bar qq $ condensates ($\langle \bar qq\rangle$)
which is  an  order parameter
of the chiral symmetry of the QCD vacuum.
In this framework a significant decrease of $\langle \bar qq\rangle$ is 
expected not only at high temperature but also  at normal nuclear
density \cite{DL91}.
Using the QCD sum rule, 
Hatsuda and Lee  theoretically predicted in-medium modification of 
 the vector mesons \cite{HL92}.
According to this model, mass decrease  at normal nuclear density
is  120 $\sim$ 180 MeV/c$^2$ for the $\omega$ and $\rho$ mesons and 
20 $\sim$ 40 MeV/c$^2$ for the $\phi$ meson. Thus the measurements
of such mesons, which are produced and decayed in a nucleus,
are of great interest. The present experiment is one of the 
several experimental efforts \cite{experiments} to investigate 
the in-medium properties of the vector mesons at 
normal nuclear-matter density.

The present experiment, KEK--PS E325, was designed to measure 
the decays of the vector mesons,
$\phi \rightarrow e^+e^-$, $\rho / \omega \rightarrow e^+e^-$
and $\phi \rightarrow K^+K^-$,
in the kinematical region where the decay probability inside 
a target nucleus was enhanced 
(0.6 $< y_{\ ee} <$ 2.2,\   0.0 $< P_{t\ ee} <$ 1.5).
The spectrometer was built at the primary beam line EP1B
in the 12GeV-PS at KEK. 
We have been taking the data since 1998, and 
this manuscript describes 
the e$^+$e$^-$ triggered data of  5.6 $\times$ 10$^7$ events collected 
in 1998
using 2.2 $\times$ 10$^{14}$ protons on the targets.
The layout of the detectors is  shown in Figure \ref{fig_spectrometer}.
The spectrometer had two electron arms and two kaon arms, which share
the dipole magnet and the tracking devices.
The electron arms covered from  $\pm$12$^{\circ}$ to $\pm$90$^{\circ}$
horizontally and $\pm$22$^{\circ}$ vertically. The kaon arms covered
from $\pm$12$^{\circ}$ to $\pm$54$^{\circ}$ horizontally and 
$\pm$6$^{\circ}$ vertically.  
Primary protons with typical  intensity of 7$\times 10^8$ Hz
were delivered to the targets at the center of the dipole magnet. 
We used three kinds of targets aligned in-line, 
carbon, polyethylene and copper,
with  interaction lengths of 0.028\%, 0.061\% and 0.020\%,
respectively.  The carbon and copper  targets were glued on 
the target supports made of paper (C$_6$H$_{12}$O$_6$)
 whose interaction length was 0.033\%. 
The combination of the intense 
beam with the thin targets were essential to keep the 
gamma conversion rate below the Dalitz decay rate. 
Typical interaction rate was as high as 1.2 MHz.

Tracking was performed with  the cylindrical drift chamber (CDC) 
and  the barrel-shaped drift chambers (BDC). 
The dipole magnet which had circular pole pieces of 880 mm in radius,
provided the field of 0.81 Tm from the center to 1600 mm in radius 
where the BDC's were located.
The acceptance of CDC was  from  $\pm$12$^{\circ}$ to $\pm$132$^{\circ}$
horizontally and $\pm$22$^{\circ}$ vertically, while BDC 
had  the same vertical coverage as CDC with a smaller horizontal coverage 
from  $\pm 7.5^{\circ}$ to $\pm 94.5 ^{\circ}$. 
The CDC consists of 10 layers of drift cells with an XX'U-VV'XX'-UXX'
 configuration
covering the radial region from r=445 mm to r=830 mm.
The BDC's have 4 layers with an  XX'UV configuration, located at
r=1600--1650 mm. 
In the X and the X' layers the direction of wires was 
 vertical and in the U and  V layers the 
wires were tilted by about $\pm$ 0.1 radian.
All the drift  cells of CDC had the same horizontal angular 
coverage of 1.5$^{\circ}$ with 
respect to the target, and the drift cells
of BDC had  the angular coverage of 0.75$^{\circ}$.
Both CDC and BDC used Argon-ethane mixed gas of 50\% and 50 \% 
at 1 atm. The total thickness of the materials from the 
target to the front of BDC was 3.1 \% radiation length.
The position resolution of 350 $\mu m$  was obtained
in the present analysis.

For the electron identification the whole region of the electron arm was 
covered by two stages of electron identification counters.
The first stage of the electron identification  was done by 
the front gas-\v Cerenkov counters (FGC) which covered from 
$\pm$12$^{\circ}$ 
to $\pm$90$^{\circ}$ horizontally and $\pm$23$^{\circ}$ vertically.
They were horizontally segmented into 13 units in each arm so that
one segment covered 6$^{\circ}$.
The second stage  consisted of three types of 
electron-identification counters,  the 
 rear gas-\v Cerenkov counters (RGC)  which covered $\pm$12$^{\circ}$ to 
$\pm$54$^{\circ}$ horizontally and $\pm$6$^{\circ}$ vertically with 7 
horizontal segments 
in each arm. These regions corresponded to the kaon-arm acceptance.
The rear lead-glass EM calorimeters (RLG) covered the same horizontal 
angle as RGCs with 12 segments in each arm, but vertically covered 
outside the kaon-arm acceptance, 
from $\pm$5$^{\circ}$ to $\pm$23$^{\circ}$. In the backward 
region where the horizontal angle was larger than 57$^{\circ}$, the second-stage
electron identification was done by the side lead-glass EM
calorimeters (SLG) which covered $\pm$57$^{\circ}$ to 
$\pm$90$^{\circ}$ horizontally and $\pm$23$^{\circ}$ vertically
 with 9 horizontal segments in each arm.

Both of the gas \v Cerenkov counters used iso-butane 
with refractive index of 1.0019. The threshold momentum 
for pions were 2.7 GeV/c. 
Both RLG and SLG consisted of  SF6W lead-glass.
 Typical energy resolution of RLG and SLG was 15\%/$\sqrt{E}$.
To suppress the fake triggers caused by pions and protons, 
we set the discriminater threshold for the gas 
\v Cerenkov counters  higher than the optimum setting 
for electrons, thus the electron efficiency was sacrificed.
For electrons with a momentum greater than 400 MeV/c,
the overall efficiencies including the trigger 
threshold and the off-line cut were 55\% for FGC,
86\% for RGC and 85\% for the calorimeters, RLG and SLG.
We achieved  the pion rejection of  6.7$\times 10^{-4}$
with a cascade operation of FGC and the EM calorimeters,
and 3.9$\times 10^{-4}$  with FGC and RGC  
for 400 MeV/c pions.

The electron trigger of three levels was adopted for the 
data accumulation. In the first level,
we selected electron-pair candidates using the coincidence 
signal of the front-stage (FGC) and the  rear-stage detectors
(SLG, RLG, RGC), by requiring the horizontal 
position matching.
To suppress electron pairs with a small opening angle,
such as from  Dalitz decays and $\gamma$ conversions,
 we required the two FGC hits to be  more than  2 segmentations 
apart.
In the second level, we required the pair to be oppositely
charged by 
using the drift chamber hits associated with the FGC hits. 
The hit positions in the chambers were obtained 
from the OR-ed signal of the X and X' layers
in the outer-most layers in CDC and in BDC.
The OR-ed signals provided an effective angular segmentation 
of 1.5$^{\circ}$ both at r=825 mm (CDC) and r=1605 mm (BDC).
The sign of a track can be roughly determined with these
segmentations together with the target position.
We eliminated the electron pair candidates which has an apparent
``$++$'' or  ``$--$'' configuration.
We also required the remained pairs 
to be  more than 12$^{\circ}$ apart at the r=825 mm position 
to suppress electron pairs with a small opening angle.
In the third level, approximate
opening angles of the pairs were calculated.
The radius of the BDC layers was almost twice larger 
than the radius  of the outer-most CDC layers, 
so that the opening angle of the 
pair at the target, $\Theta _{open}$, could be approximated by 
$\Theta _{open} = 2 \times \Theta _{cdc} - \Theta _{bdc}$,
where $\Theta _{cdc}$ was the opening angle of the 
pair at the outer-most CDC layer and 
$\Theta _{bdc}$ at the BDC layer with respect to the target position.
We required the $\Theta _{open}$ to be in the range 
from 50$^{\circ}$ to 150$^{\circ}$ in the trigger.
The typical trigger rates were 450, 330 and 280 Hz in the first,
second and third level triggers, respectively.
We kept the live time of the data acquisition to be around 60\%.

To evaluate the performance of the spectrometer,
the mass resolution was examined for
 the observed peaks of 
$\Lambda$ $\to$ $p +\pi ^-$ and K$_s$ $\to$ $\pi ^+ +\pi ^-$ 
decays  as shown in Figure \ref{fig_massres}a and 
\ref{fig_massres}b.
For the $\Lambda$ peak we obtained the centroid at 
1115.5  MeV/c$^2$
(known to be 1115.7 MeV/c$^2$) with the Gaussian resolution of  
1.8 $\pm$ 0.1 MeV/c$^2$, 
and for K$_s$ 493.9 MeV/c$^2$ 
(known to be 497.7 MeV/c$^2$) and 
3.6 $\pm$ 0.6 MeV/c$^2$, respectively. The observed peak positions 
and the widths give the systematic uncertainty of the 
mass scale and the mass resolution of the present analysis.
The results were compared to the  Monte Carlo simulations 
in which we took the chamber resolution and the multiple 
scattering into account. 
The observed widths were well reproduced by the simulation 
(1.9 MeV/c$^2$ for $\Lambda$ and 
3.5 MeV/c$^2$ for K$_s$), and 
the energy scale uncertainty for the $\omega$ and $\phi$ meson was 
estimated to be 4 MeV/c$^2$ and 7 MeV/c$^2$  and 
the mass resolution to be 9.6 MeV/c$^2$ and 12.0 MeV/c$^2$,
respectively. 

Figure \ref{fig_eemass} shows the e$^+$e$^-$ invariant mass spectra;
(a) for the carbon and polyethylene targets (light nuclear targets) and 
(b) for the copper target (heavy nuclear target). 
These histograms contain the events when the  electron and the positron 
are detected in the different arms, so that the low mass part of the 
spectra is largely suppressed.
Although the clear peaks of the $\omega$ meson
 decaying into e$^+$e$^-$ is visible in the spectra,
a significant shape difference is observed between 
the light  nuclear targets and the copper target and
an excess at the low mass side of the $\omega$ peak can be seen.

We have tried to reproduce 
the mass  shape of the obtained histograms
with   the combinatorial background and the known hadronic sources.
The origins of the combinatorial background were the pairs which 
were picked up from two independent Dalitz decays or $\gamma$ conversions,
and the pairs like e$^- \pi ^+$ or e$^+ \pi ^-$ due to 
the mis-identification. 
The remaining e$^- \pi ^+$ and e$^+ \pi ^-$ background 
was estimated to be about 13\% in the spectra
and the contaminations like $\pi ^+ \pi ^-$ to be  negligibly
small. 
The distribution of the combinatorial background was obtained 
from the event mixing method.
As the known hadronic sources, $\rho \to e^+ e^-$, 
$\omega \to e^+ e^-$,
$\phi \to e^+ e^-$,
$\eta \to e^+ e^- \gamma$ and 
$\omega \to e^+ e^- \pi ^0$  were considered. 
The Dalitz decay, $\pi ^0 \to e^+ e^- \gamma$, is negligible in the 
mass acceptance of the present data.
The shapes of the $e^+ e^-$ invariant mass spectra from the Dalitz
decays, $\eta \to e^+ e^- \gamma$ and $\omega \to e^+ e^- \pi ^0$,
 were taken from the reference \cite{Faessler}.
The mass shape of the $\rho$, $\omega$ and  $\phi$  mesons was 
given as the Breit-Weigner function 
with the natural width 150 MeV/c$^2$, 8.41 MeV/c$^2$ 
and 4.43 MeV/c$^2$, respectively.
The Breit-Weigner functions were smeared with the estimated mass 
resolution of 9.6 MeV/c$^2$ for the $\omega$ and 12.0 MeV/c$^2$
for the $\phi$ meson.
The present experiment used the targets with radiation length of 
less than 0.3\% so that the radiative tail and the multiple 
scattering due to the target thickness were negligible.
We have assumed that the production cross section of $\rho$ 
is equal to  that  of $\omega$ \cite{Blobel:1974wr}.
To obtain the mass shape of the known sources in the observed 
spectra, we evaluated the experimental mass acceptance 
using the particle distributions obtained by
the nuclear cascade code, JAM \cite{jam}.

The relative abundances of the known sources  and 
the combinatorial background  were obtained through the fit
with four  parameters,
the amplitudes  of  $\rho$/$\omega$ $\to e^+ e^- $, 
$\phi$ $\to e^+ e^- $, 
$\eta$ $\to \gamma e^+ e^- $ and the combinatorial background. 
The amplitude of the other source, 
$\omega$ $\to \pi ^0 e^+ e^-$, was given from the
branting ratio. The best fits are over plotted in 
Figure \ref{fig_eemass}.  The contribution of 
$\eta$ $\to \gamma e^+ e^- $ turned out to be negligible.
As the result of the fits, 
we found 75.5 $\pm$ 9.0 $\omega$ mesons  and 
7.4 $\pm$ 5.8 $\phi$ mesons  from the light target 
and  20.0 $\pm$ 4.8 $\omega$ mesons and 
5.2 $\pm$ 2.7 $\phi$ mesons 
from the copper  target.
The invariant mass shapes were well reproduced 
except the mass region below the omega peak.

To evaluate the excess in the mass region from 550 MeV/c$^2$
to  750 MeV/c$^2$,  we fit the histograms again excluding 
this mass region. 
The excess was  estimated by subtracting the amplitude
of the fit function from the data. 
The number of excess of the light target is 
19.6 $\pm$ 11.7 and that of the copper target is 
29.5 $\pm$ 8.7.
The excess is statistically significant for the copper target data.
The ratios to the amplitude of the omega peak 
are 0.26 $\pm$ 0.16 for the light target and 
 1.48 $\pm$ 0.56  for the copper target.
The difference between the two cases should be 
originated from the difference of the nuclear size.
The natural explanation of the shape change is that 
the mass modification of $\rho$/$\omega$ mesons 
takes place inside a nucleus. Although the mass shape 
of the modified meson is difficult to predict,
it should be noted that the 
excess in the  copper target data 
is visible in the mass range about 200 MeV below the 
$\omega$ peak. This range is consistent with the 
expected shift predicted by Hatsuda and Lee \cite{HL92}.

In summary we have observed a signature of in-medium mass 
modification of $\rho$/$\omega$ meson. 
This is the first observation of the leptonic in-medium decay of
the vector mesons at normal nuclear-matter density.\\

\noindent
{\bf Acknowledgements }
We would like to thank all the staff of KEK PS,
especially for the helpful support kept by the beam 
channel group.
This work was partly supported by 
Japan Society for the Promotion of Science and 
a Grant-in-Aid for Scientific Research of the Japan Ministry
of Education, Science and Culture (Monbusho).

%
%

 \begin{figure}
  \caption{A schematic view of the experimental setup of the E325 
 spectrometer; (a) for the top view and (b) for the side view. The figure of 
 the side view shows the cross section along the center 
 of the kaon arm (see text).}
 \label{fig_spectrometer}
 \end{figure}

\begin{figure}
 \caption{Invariant mass spectrum of p$\pi ^-$ (a)
   and $\pi ^+ \pi ^-$ (b).
   The lines are 
   the best fit results applying  the Gaussian with a  linear 
   background.  The vertex position of p$\pi ^-$ pairs were required to 
   be more than 20 mm apart from the target
   and of $\pi ^+ \pi ^-$ pairs more than  10mm.}
 \label{fig_massres}
 \end{figure}

 \begin{figure}
 \caption{Invariant mass spectrum of e$^+$e$^-$ pair;
   (a) for the carbon and polyethylene targets and (b) for the
   copper target.
   The solid lines show  the best fit results of  
   the known hadronic sources with the combinatorial background.
   The dotted lines are the contribution from 
  $\rho$, $\omega$ and $\phi$ decays.
   The dashed lines are for $\omega$ $\to$  $\pi^0$ e$^+$e$^-$  decays 
   and the  dot-dashed  lines are for the combinatorial background.}
 \label{fig_eemass}
 \end{figure}

%
%
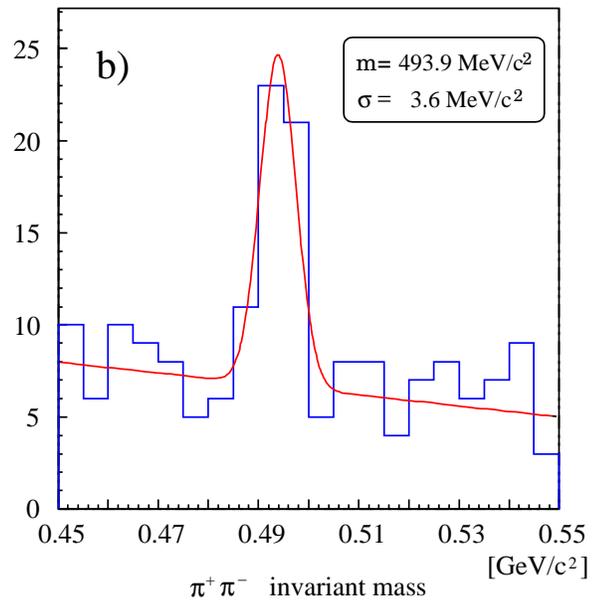
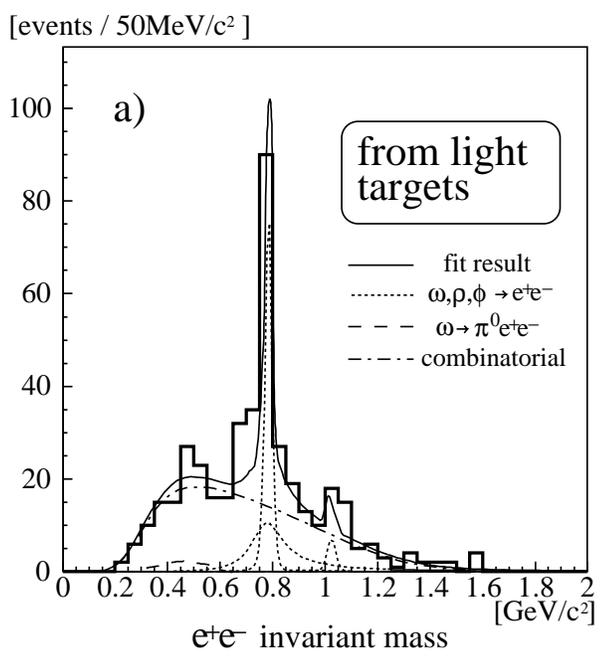
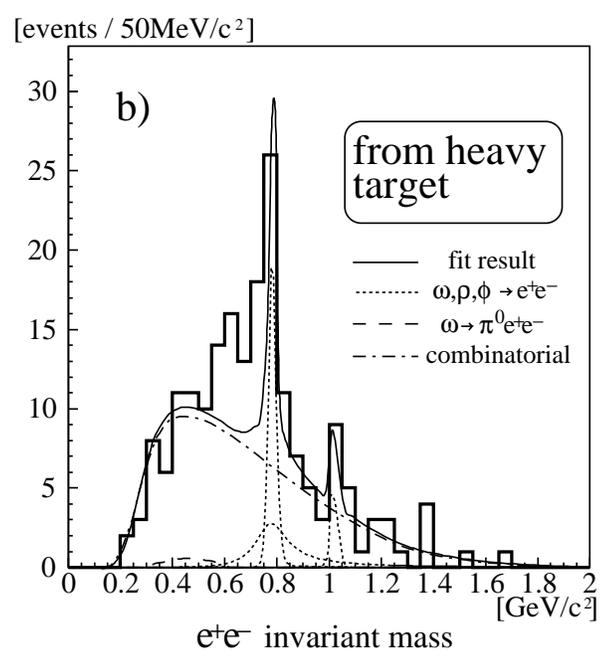
\end{document}